# Design, Development, and Verification of the Planck Low Frequency Instrument 70 GHz Front-End and Back-End Modules

**J. Varis**[a*], **N.J. Hughes**[b], **M. Laaninen**[c], **V.-H. Kilpiä**[b], **P. Jukkala**[b], **J. Tuovinen**[a], **S. Ovaska**[c], **P. Sjöman**[b], **P. Kangaslahti**[d], **T. Gaier**[d], **R. Hoyland**[e], **P. Meinhold**[f], **A. Mennella**[g], **M. Bersanelli**[g], **R.C. Butler**[h], **F. Cuttaia**[h], **E. Franceschi**[h], **R. Leonardi**[f], **P. Leutenegger**[j], **M. Malaspina**[h], **N. Mandolesi**[h], **M. Miccolis**[i], **T. Poutanen**[j, k, l], **H. Kurki-Suonio**[j, k, l], **M. Sandri**[h], **L. Stringhetti**[h], **L. Terenzi**[h], **M. Tomasi**[g] and **L. Valenziano**[g]

[a] *MilliLab, VTT Technical Research Centre of Finland,*
*P.O. Box 1000, FI-02044 VTT, Finland*
*E-mail:* Jussi.Varis@vtt.fi

[b] *DA-Design Oy,*
*Keskuskatu 29, FI-31600 Jokioinen, Finland*

[c] *Ylinen Electronics Oy,*
*Teollisuustie 9A, FI-02700 Kauniainen, Finland*

[d] *Jet Propulsion laboratory,*
*Pasadena, CA, USA*

[e] *Instituto de Astrofisica de Canarias,*
*La Laguna, Tenerife, Spain*

[f] *University of California, Santa Barbara, Department of Physics,*
*Santa Barbara, CA, USA*

[g] *Università Degli studi di Milano, Dipartimento di Fisica,*
*Milano, Italy*

[h] *INAF/IASF,*
*Via P. Gobetti 101, I-40129 Bologna, Italy*

[i] *Thales Alenia Space Italia S.p.A., IUEL – Scientific Instruments,*
*S.S. Padana Superiore 290, I-20090 Vimodrone (Mi), Italy*

[j] *University of Helsinki, Department of Physics,*
*P.O. Box 64, FI-00014 University of Helsinki, Finland*

[k] *Helsinki Institute of Physics,*
*P.O. Box 64, FI-00014 University of Helsinki, Finland*

[l] *Metsähovi Radio Observatory, Helsinki University of Technology,*
*Metsähovintie 114, FI-02540 Kylmälä, Finland*

ABSTRACT: 70 GHz radiometer front-end and back-end modules for the Low Frequency Instrument of the European Space Agency's Planck Mission were built and tested. The operating principles and the design details of the mechanical structures are described along with the key InP MMIC low noise amplifiers and phase switches of the units. The units were tested in specially designed cryogenic vacuum chambers capable of producing the operating conditions required for Planck radiometers, specifically, a physical temperature of 20 K for the front-end modules, 300 K for the back-end modules and 4 K for the reference signal sources. Test results of the low noise amplifiers and phase switches, the front and back-end modules, and the combined results of both modules are discussed.

---

[*] Corresponding author.

At 70 GHz frequency, the system noise temperature of the front and back end is 28 K; the effective bandwidth 16 GHz, and the 1/f spectrum knee frequency is 38 mHz. The test results indicate state-of-the-art performance at 70 GHz frequency and fulfil the Planck performance requirements.

**Contents**

**1. Introduction**

Two multi-pixel and ultra-sensitive instruments are aboard the European Space Agency's (ESA) Planck Mission to determine the anisotropy of the Cosmic Microwave Background (CMB) radiation. The mission was launched in 2009 [1]. The Low Frequency Instrument (LFI) has channels at 30, 44 and 70 GHz with 20% bandwidth. The number of dual polarised receiving horn antennas is 2, 3 and 6, respectively. Since each horn has two orthogonal polarisations, the number of receivers is twice the number of horns. The High Frequency Instrument (HFI) has six channels and a total of 52 receivers covering the 100-850 GHz range. At 70 GHz, an angular resolution of about 14' and a sensitivity of 11 µK per sky map pixel will be reached.

    This paper describes the development of the 70 GHz receivers' front-end modules (FEMs) and back-end modules (BEMs) in Finland. The development was led by MilliLab in collaboration with Ylinen Electronics. The model philosophy for developing the 70 GHz receiver unit was divided into four steps: the Prototype Demonstrator (PTD, [2]), the Elegant Breadboard Model (EBB, [3, 4]), the Engineering Model (EM), and the Protoflight Model (PFM). The six PFM FEMs and BEMs are the actual flight units in the Planck spacecraft, and their design and performance will be described in detail here. During the development stages, the mechanical structure and the electronic circuits underwent many changes to reach the final modular design (figure 1), which fulfils the mass and power requirements. In addition from design point of view, Flight Spares (FS) were built identical to the PFMs.

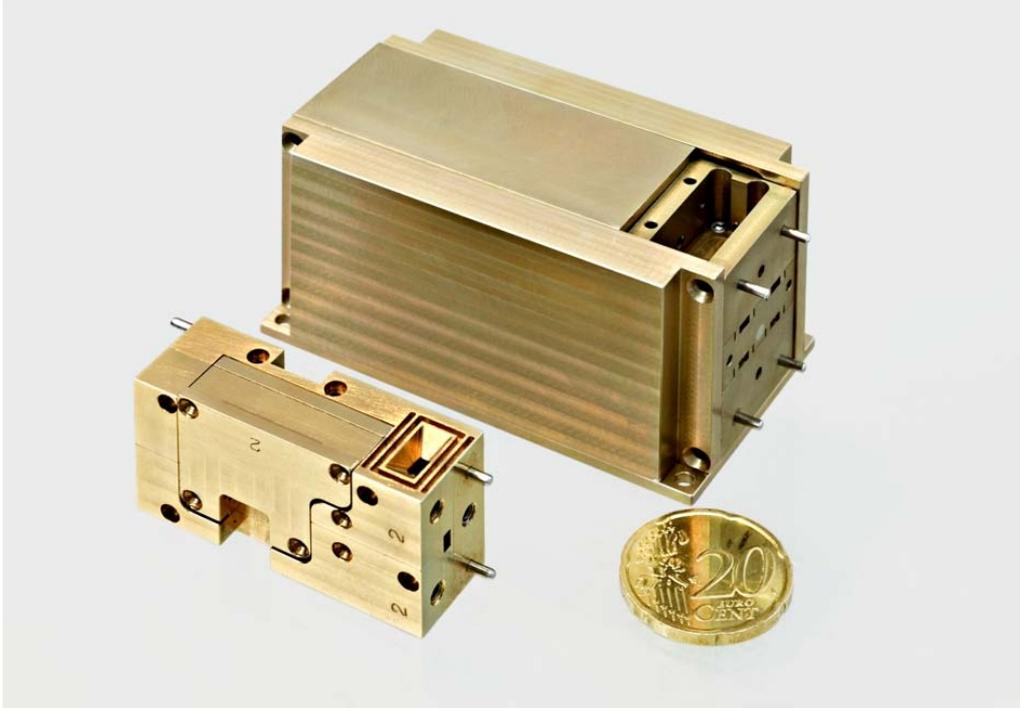

**Figure 1.** A photograph of the Planck LFI 70 GHz FEM (half of one module) and BEM (a full module).

In this paper, different parts of the receiver are described in the order to understand how the signal passes from the sky through the RF circuitry. The passive components are discussed elsewhere [5-7]. First, the basic receiver concept is described in section 2. The FEM, the heart of the receiver, operating at 20 K and its key High Electron Mobility Transistor (HEMT) Low Noise Amplifiers (LNAs) are detailed in section 3, along with the Heterojunction Bipolar Transistor (HBT) PIN phase switches. Section 4 discusses the BEM operating at 300 K. Finally, a short description of the cryogenic test facilities are described in section 5 and the measured performance of the receivers is presented in section 6.

## 2. The concept of the Planck LFI receiver

In the LFI receivers, the FEMs are cooled to 20 K while the BEMs are kept at 300 K. The LFI architecture is based on direct detection, where the signal is first amplified to approximately 60 dB, and then detected by a diode. A DC-voltage proportional to the incoming noise power is therefore obtained without any down conversion or intermediate frequencies. Cooling is used to reduce the amplifier noise in the front end and to generally improve the signal to noise ratio of the radiometer. A clear advantage of this scheme is the inherent simplicity of the radiometer, though the extreme stability required for the amplification chain is clearly a disadvantage. The need for stability required the use of continuous comparison architecture in the receiver's design. The time scale of the stability of the receiver is driven by the 1 rpm rotation speed of the Planck spacecraft (corresponding to a frequency of 17 mHz). During this single rotation period, the system noise and gain should not vary significantly. This means that all the RF chains have to be designed to have a very low 1/f noise.

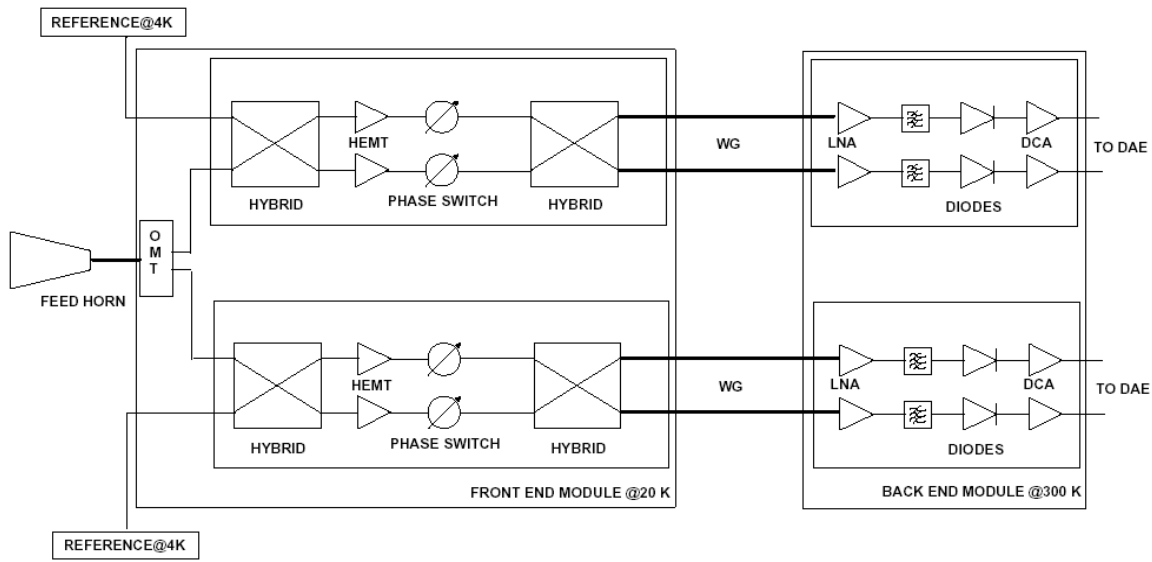

**Figure 2.** A block diagram of the baseline Planck LFI radiometer.

The general schematic of the radiometer is shown in figure 2. The CMB signal (or 'sky' signal) is received with a tapered horn antenna and divided into two orthogonal, linearly polarised components using an OrthoMode Transducer (OMT). Each component is fed to an identical receiver that continuously compares the one polarisation to a 4 K microwave reference source ('ref'). In order to do that both signals are connected to the inputs of a waveguide hybrid coupler (the so-called 'magic-T'), which feeds one of the input signals from both coupler outputs with the same phase and the other signal with a 180º phase shift between the outputs. The two input signals are thus combined in phase ('sky'+'ref') and in anti-phase ('sky'-'ref'). Next, the magic-T output signals are amplified by HEMT LNAs, phase-switched through 0/180º phase switches and re-combined by a second magic-T coupler. The second hybrid is necessary to re-separate the one polarisation sky signal and the reference. In the LFI, the phase switch is modulated by the Data Acquisition Electronics (DAE) at 8192 Hz, producing alternating signals at 4096 Hz. These signals are then conveyed via waveguides to the back-end module, where filtering, further amplification and detection take place. After detection, the four BEM output signals (two single polarisations and two references) are acquired by the DAE. The DAE samples the DC voltages produced by the BEM at the 8192 Hz frequency.

In the FEM, both the sky and reference signals go through the two almost identical amplifier chains and are subjected to the same gain variations. Provided that the gain profiles of the two chains are very similar and there is high isolation between the RF chains during different phase switch states, the 1/f noise contribution of the amplifiers can be greatly reduced subtracting the output signals from each other. In the BEM, the amplifier-detector chains are directly amplified without switching, and therefore the 1/f noise cannot be reduced considering that the switching technique is at RF level. The only option involves switching the signals between the BEM inputs rapidly as described above. From the digitised data produced by the DAE, sky-only and reference-only data streams are generated and then subtracted from each other to produce a differenced stream in which most RF chain noise has been removed. This subtraction takes place off line on the ground and not onboard the spacecraft. The operation of the Planck LFI receivers is discussed in detail in [8, 9].

## 3. Front-End Module

### 3.1 The FEM circuit

Since each FEM contains two identical continuous comparison receivers for each polarisation, it is sufficient to describe the structure of a single receiver ('half FEM').

In the half FEM module, the horn antenna, the OMT, and the 4 K reference source are external. The reference source is a microwave load [6] attached to the 4K cooler of the spacecraft. This load is viewed by a rectangular horn antenna, which was machined as part of the body of the half FEM receiver. Furthermore, the two magic-T couplers and waveguides to carry the signals (figure 3 a) were machined in the body of the half FEM. The 20% bandwidth requirement with centre frequency 70 GHz means 14 GHz (63-77 GHz instantaneous bandwidth). This is covered with WR-12 waveguides (E-band, 60-90 GHz). The half FEM weight 42 grams is below the Planck requirement of 50 grams.

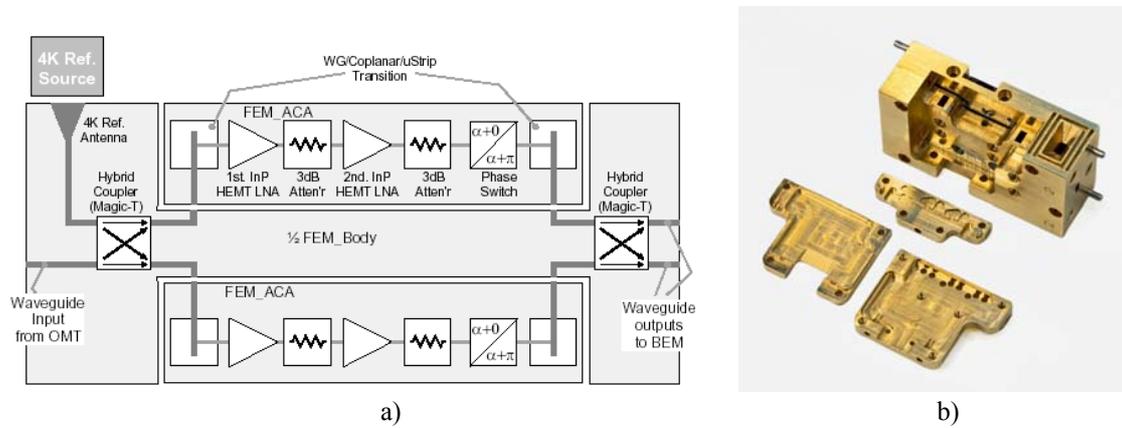

a) b)

**Figure 3.** a) A block diagram of a half of a front-end module. b) A photograph of the mechanical structure of an exposed half FEM. The parts of an amplifier chain assembly are shown from the front. The horn antenna viewing the 4 K reference load is shown on top and to the right of the half FEM structure.

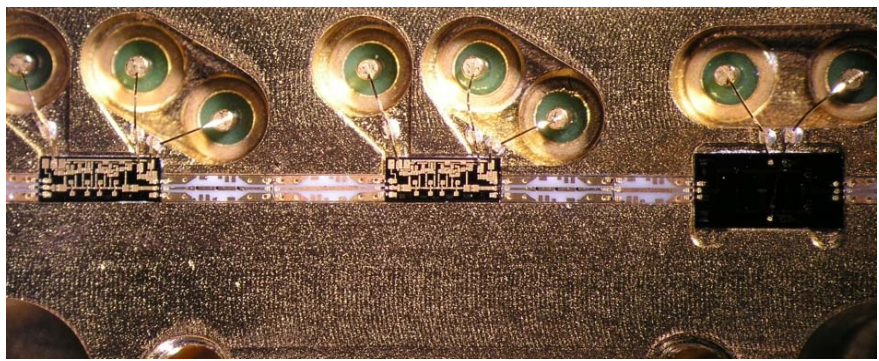

**Figure 4.** A detail of a FEM amplifier chain assembly showing the two LNAs (left and middle) and the phase shifter (right) forming the amplifier chain.

The two identical Amplifier Chain Assemblies (FEM_ACA), which amplify the combined signals in phase and anti-phase from the coupler outputs, were designed to be removable and exchangeable (figure 3 b). Within these assemblies the signals are converted from waveguide form to microstrip transmission line and then to coplanar transmission line in order to be compatible with the low noise amplifier input connections (figure 4).

Each FEM_ACA has two HEMT LNAs and one phase switch with auxiliary passive components contained in the FEM_ACA casing. The biases of the amplifiers can be controlled from the data acquisition electronics unit to obtain the best possible amplification with the lowest noise operation. The DAE also controls the biases of the phase switch and generates the signal necessary to alternate the switch state from 0° to 180°. The output signals of the FEM_ACAs are then conveyed, via further coplanar to waveguide transitions, to the output coupler for the re-combination. Perfectly equalised between the two amplifier chains, the signals from the second coupler outputs will correspond to amplified versions of the sky and the 4 K ref input signals.

### 3.2 Low noise amplifiers

The 70 GHz LFI receivers are based on Monolithic Microwave Integrated Circuit (MMIC) semiconductors, and the key components are the LNAs. Because of the very low noise performance requirements, the MMIC LNAs were made using an Indium Phosphide (InP) HEMT technology. The InP technology enables very low power operation in addition to low noise. This is essential for the FEM operated at 20 K. The overall gain of the 70 GHz receiver is about 60 dB, where 35 dB are obtained from the FEM. The power consumption of a FEM should not exceed 24 mW at 20 K.

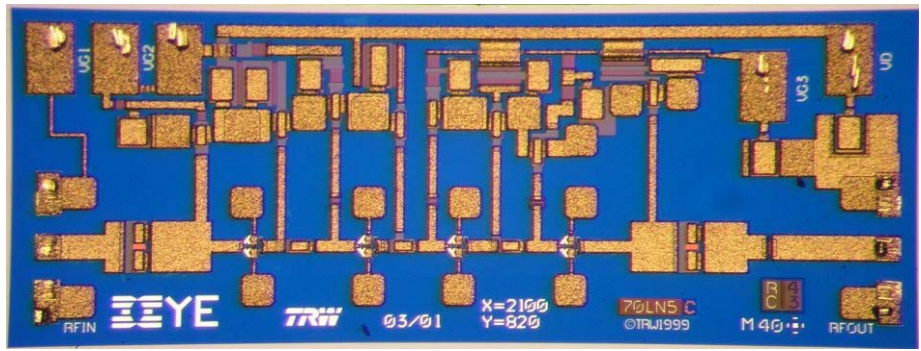

**Figure 5.** A four-stage InP MMIC low noise amplifier used in the 70 GHz Protoflight Models. The size of the MMIC is 2.1 mm x 0.8 mm.

Several MMIC foundries were evaluated for manufacturing the MMICs for the Planck LFI. For the final manufacturing, the NGST (previously TRW) InP HEMT process with a nominal 0.1 μm gate length was selected. This process has via-holes and backside metallisation capability, which enables the use of microstrip transmission lines. Figure 5 shows a photograph of a four-stage microstrip design from a NGST wafer. Both coplanar and microstrip designs were studied, but in both architectures the performance was highly similar. The amplifier designs are discussed in more detail in [10, 11].

Figure 6 a) shows the on-wafer room temperature measurement of gain and noise figure of one of the four-stage amplifiers. Its gain exceeds 25 dB, and the noise figure is 2.5 dB at 70 GHz. The uncertainty of both the gain and the noise figure is ±0.1 dB, which is a worst case estimate based on information provided by the test equipment manufacturer. The power consumption is 17.5 mW in this case, which is rather high. When used in the amplifier chain assembly at 20 K with another twin LNA and the phase switch, the gain is better than 35 dB with a very low noise contribution (figure 6 b). The noise temperature is 24.7 K (0.3 dB noise figure) at 70 GHz, and on average, 24.8 K in the 63-77 GHz range. Also, the power consumption is only 4.6 mW for both LNAs combined.

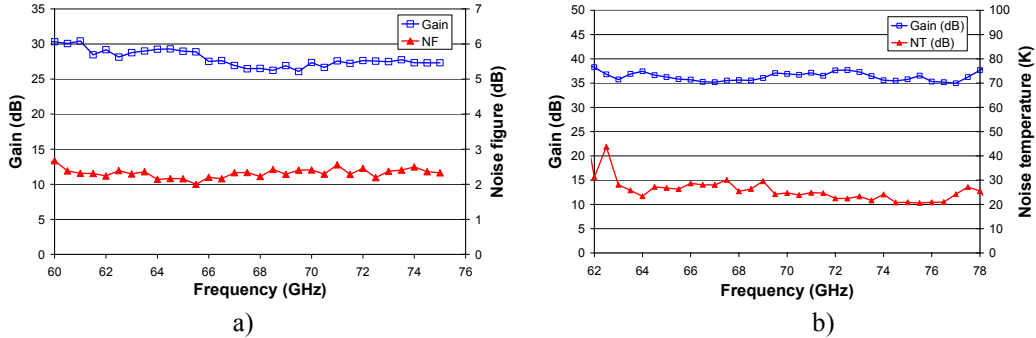

**Figure 6.** a) Measured gain and noise figure of a four-stage LNA at room temperature. b) Gain and noise temperature of a amplifier chain assembly cooled at 20 K.

### 3.3 Phase switches

As stated in Section 2, the LFI receivers utilise the continuous comparison technique to reduce 1/f noise in the amplifier chains, especially due to the HEMT LNAs gain instability. This is usually characterised by the 1/f knee frequency, which is the point where the 1/f noise spectral density is equal to the white noise level. However, this technique does not eliminate the 1/f noise of the phase switches, because they are not operated independently. For this reason, it is imperative to use phase switches with a 1/f noise contribution as low as possible. The 17 mHz value resulting from the spacecraft rotation is practically impossible to reach with MMIC devices and therefore unrealistic for a radiometer as a whole. However, Seiffert et al. [8] have discussed that this requirement for the radiometers can be addressed through computational methods such as applying suitable destriping and map-making algorithms. So the final Planck requirement for the RF chain was relaxed to 50 mHz.

After evaluating various processes, the NGSTs InP HBT PIN diode process was selected for manufacturing the phase switches. The InP HBT PIN diodes are known for their very low 1/f contribution. A photograph of the 0°/180° phase switch device is shown in figure 7. The design is based on two hybrid rings back-to-back with two diodes (12 μm diodes) to control the output [12]. During the 0° state, the first diode is conducting and the second non-conducting. The opposite occurs during the 180° state. At 20 K, the transmission loss ($S_{21}$) of the device is less than 3 dB in the overall 63-77 GHz band (figure 8 a). The frequency responses of the two phase switch states are practically identical and so the amplitude balance (the difference of amplitude between the two states) is excellent (figure 8 b). In the 63-77 GHz range, the balance varied ±0.13 dB. The phase shift between the two switch states was an almost ideal 180°: 181.6°±0.9°. The power consumption of a single diode is approximately 0.8 mW during the conduction.

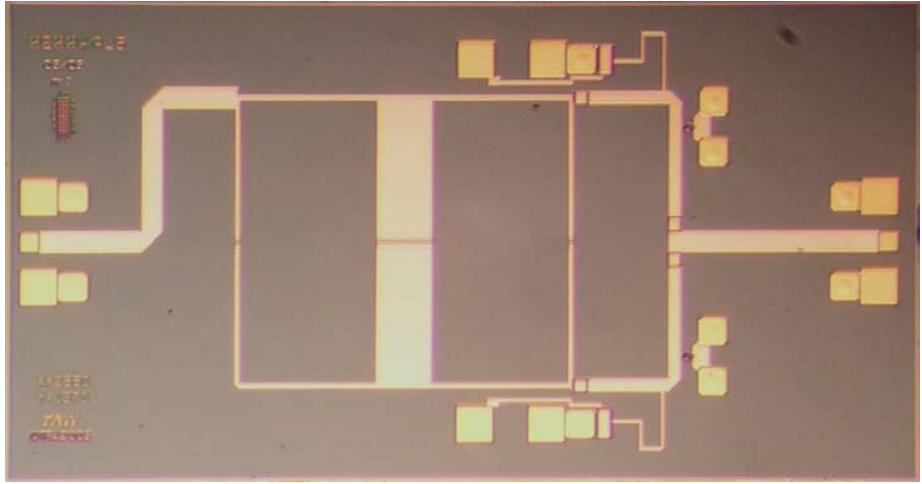

**Figure 7.** An InP MMIC phase switch used in the 70 GHz amplifier chain assemblies. The size of the MMIC is 2.5 mm x 1.4 mm.

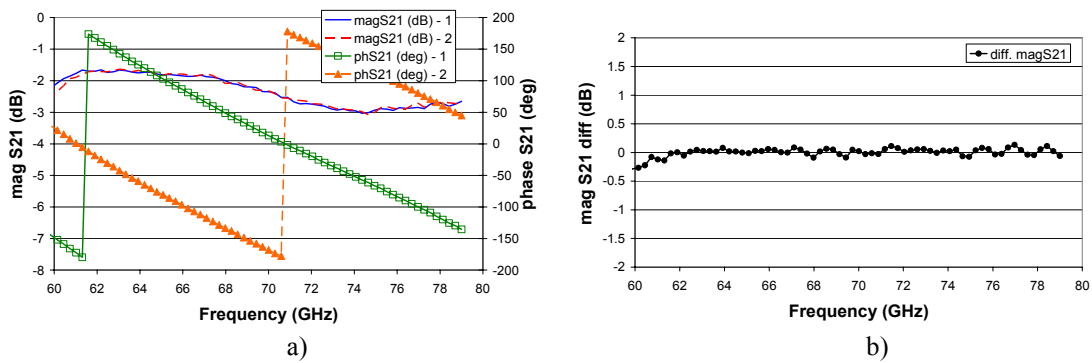

**Figure 8.** a) Measured transmission loss ($S_{21}$) magnitudes and phases for two phase switch states. b) Amplitude balance between the two phase switch states. The physical temperature of the phase switch was 20 K.

## 4. Back-End Module

The FEM output signals are conveyed by waveguides (WR-12) to the back-end modules (see figure 2), housed with the Data Acquisition Electronics (DAE) assembly. To maintain compatibility with the FEM, each BEM accommodates four receiver channels from the four waveguide outputs.

In each channel within the BEM, the signal passes through a waveguide band pass filter (figure 9 a), which limits the signal bandwidth to the target range of 63-77 GHz. Figure 10 shows one of the waveguide filter frequency responses. The -3 dB bandwidth range is from 62-81 GHz, which in this case is slightly wider than required. The transmission loss is less than 0.5 dB ± 0.1 dB. Here the measurement uncertainty is a worst case estimate based on information provided by the VNA manufacturer. The accuracy of the frequency measurement is in the sub-megahertz range according to the manufacturer. The frequency step used in the measurement was 0.125 GHz giving a rough estimate for the bandwidth measurement uncertainty (~± 0.06 GHz).

The signal connected from the waveguide-to-coplanar transition is amplified by a single MMIC LNA and finally detected by a diode. The filters and the amplifier-detector assemblies with the transitions were designed as separate modules to allow better freedom in their testing (figure 9 b). The filter unit contains filters for two channels to accommodate the outputs of a half FEM. Similarly the amplifier-detector assembly contains LNAs and diodes for two channels. Therefore, each BEM has two filter units and amplifier-detector assemblies installed to the main BEM casing. The casing also contains DC amplifiers, bias protection circuits and connectors. The BEM power consumption is less than or equal to 604 mW. The BEM weighs 154 grams fulfilling the Planck requirement of 164 grams.

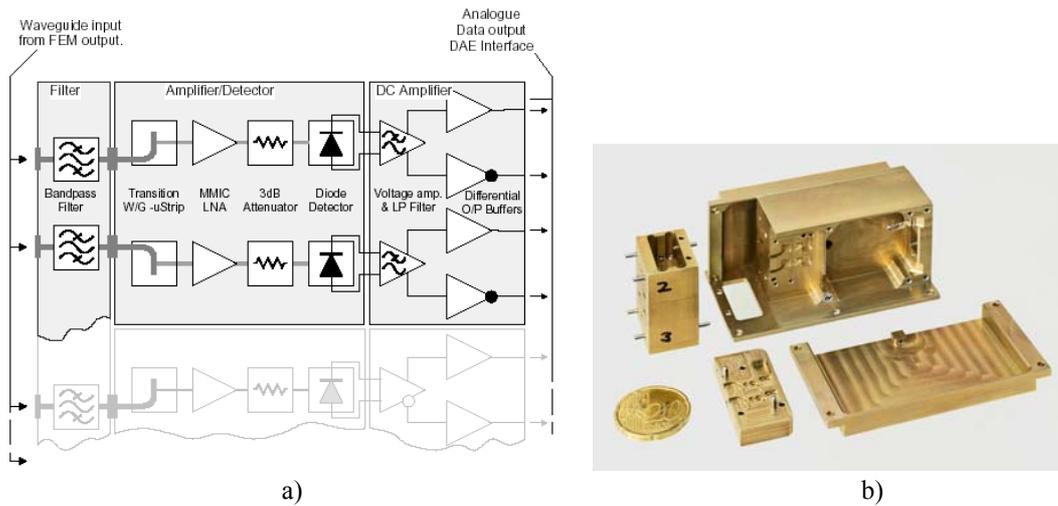

a) b)

**Figure 9.** a) A block diagram showing half of a back-end module. b) A photograph of the mechanical structure of the 70 GHz BEM. The waveguide band pass filter is shown on the left. The front and middle section houses the microwave amplifier-detector assembly. The DC amplifier, protection circuits and connectors are placed within the cover plate (at the front and to the right) and in the cavity on the right.

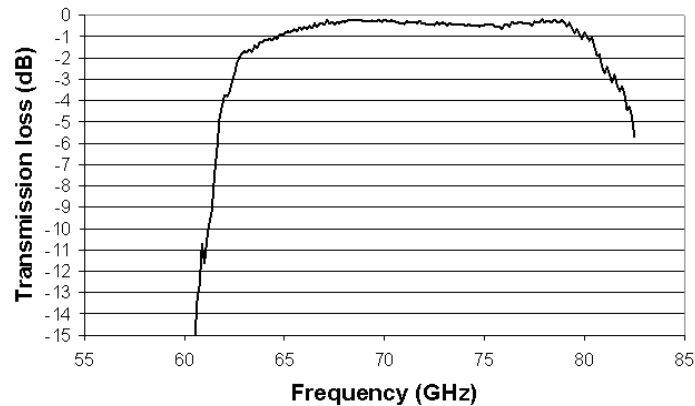

**Figure 10.** A typical BEM waveguide filter frequency response.

The amplifier-detector units use the same type of MMIC amplifiers as the FEMs. The detectors are commercially available as zero-bias Gallium Arsenide (GaAs) Schottky diodes. These are biased in their "squared-law" region providing an output voltage proportional to the input microwave power.

Further DC amplification is necessary to reach the DAE input range. This output signal is differential to help to shield against electromagnetic interference from outside and to preserve grounding integrity. Furthermore the DAE provides the power supply to the BEM.

## 5. Cryogenic testing

A custom-built vacuum chamber was constructed to house two full radiometers (figure 11, see also [13]). The dimensions of the chamber were 1.6 m x 1.0 m x 0.3 m. To reach the cryogenic temperatures necessary for the FEMs and the 4K ref and sky loads, the chamber was integrated with commercially available 4 K and 20 K closed-cycle helium coolers. The FEM bodies were mechanically supported close to the 20 K cooler cold plate with the metal support connected directly to the plate. The sky and 4K ref loads were connected to the 4 K cooler providing by heating elements the correct load temperatures. Sensors were also used to monitor the temperatures of the loads and the FEMs. In order to simulate the space environment the FEMs and the loads were also contained within a closed box radiation shield connected to the intermediate stage of the 20 K cooler. The temperature of the shield was approximately 70 K. During the tests, the loads reached a minimum temperature of about 7 K and the FEM body about 22 K. The temperature stability of the measurements was approximately ±10 mK.

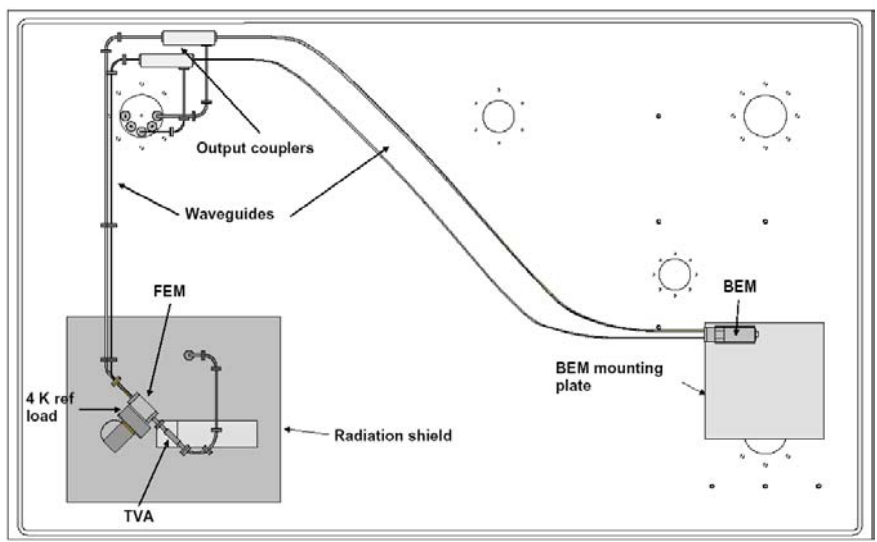

**Figure 11.** A schematic of the large cryogenic vacuum chamber used for testing the 70 GHz Protoflight Model front-end and back-end modulles combined with Planck representative waveguides and signal sources.

The 4K ref loads provided by IASF-Bologna from Italy [14] were radiative noise sources directly facing the horn antennas of the FEMs. The sky loads were Thermal Vane Attenuators (TVA) built in-house, essentially waveguides with temperature-controlled attenuating elements.

The TVAs were connected to the FEM inputs via short stainless steel waveguide sections (no horn antennas or OMTs). The TVAs were two-port devices allowing noise or CW signals to be injected as an alternative stimulus to FEM.

The FEMs were connected to the back-ends with in-house built copper waveguides (WR-12) containing also stainless steel sections for temperature isolation. In addition, the waveguides had couplers to enable only FEM or only BEM measurements without disassembling the radiometer. The waveguides' lenghts were 1.5 m, similar to which is used in the Planck spacecraft. However, the flight condition temperature gradient in the waveguides could not be duplicated in the test setup. The attenuation of a single waveguide was about 7 dB. The BEMs were mounted on a thermally isolated plate with Peltier effect heater/coolers to maintain the BEMs at a constant temperature of 300 K.

The various biases for the FEMs and BEMs are provided by the DAE, which also controls the phase switch modulation and handles the data sampling. In the tests described here, the various biases were provided with separate analogue power supplies, and the phase switch control and data sampling were performed with a custom-built data acquisition system.

The BEM frequency responses (BEM output voltage vs. stimulus frequency and amplitude) and dynamic ranges (BEM output voltage vs. stimulus power) were characterized at room temperature before the assembly of the radiometers. In these measurements, a Vector Network Analyser (VNA) was used to provide a CW stimulus to the BEM. A coupler and a power meter between the VNA and the BEM were used to determine the input microwave power to the BEM. Also, a tunable attenuator was used to adjust the input power. The BEM output voltages were measured using a digital volt meter.

During the cryogenic measurements (FEMs cooled to 20K and BEMs in 300 K), the gains and isolations of the FEMs were first measured using the VNA. In these measurements, the stimulus from the VNA was injected through the TVAs to the FEMs and the response was measured from the outputs of the couplers located in the middle of long connecting waveguides. The effect of the sky loads and waveguide losses was de-embedded from the measurements.

In the FEM and BEM combined tests, the frequency responses of the radiometers were first checked to confirm that the BEM waveguide filters cut the radiometer bandwidth as expected. The bench test was very similar to the BEM frequency response one, except the CW stimulus was fed to the FEM input and the radiometer was operating in the cryogenic conditions. The final test of the radiometers was performed while simulating the flight conditions as accurately as possible. To perform this test the 4K ref and sky loads were used as the signal sources. The data acquisition system generated the signal to modulate the phase switches and it sampled the output voltages of the BEMs at the specified 8192 Hz frequency to produce the alternating sky-ref data streams. From the difference of the time series (done off line by a computer program) , the noise spectra of the radiometers were determined using the discrete Fourier transform. From the spectra, the spectral density of the white noise and 1/f noise of the radiometers were determined. The radiometer noise temperatures were measured using the standard Y-factor technique. To perform the test the temperature of the sky load was increased in suitable temperature steps, the values typically between 5 and 30 K, reference load kept constant at the minimum temperature and the radiometer output voltages measured. By a linear curve fit, the noise temperature was determined.

# 6. Performance

## 6.1 FEMs

For the LNA selection of the PFM FEMs, nine different wafers from various NGST processing runs were evaluated. Only the LNAs with the best performance were assembled as the first stage amplifiers in the FEM_ACAs. Amplifiers from the other wafers were used in the second stage of the FEM_ACAs and in the BEMs. For the phase switch selection, four different wafers from NGST were evaluated. Although about 150 FEM_ACAs were built and tested for their gain and noise performance with a variety of bias points, only 24 for the six PFM FEMs were selected. The criteria used for the selection of LNAs are gain and noise temperature performance, but the matching of gain profiles between pairs of FEM_ACAs is important. This is necessary to provide good isolation between the two channels of each polarisation. Poor channel isolation would ruin the FEM noise reduction scheme discussed in section 2. The Planck requirement for isolation is 13 dB. In average, the gains of the selected amplifier chain assemblies ranged from 34-39 dB and noise temperatures from 22-35 K.

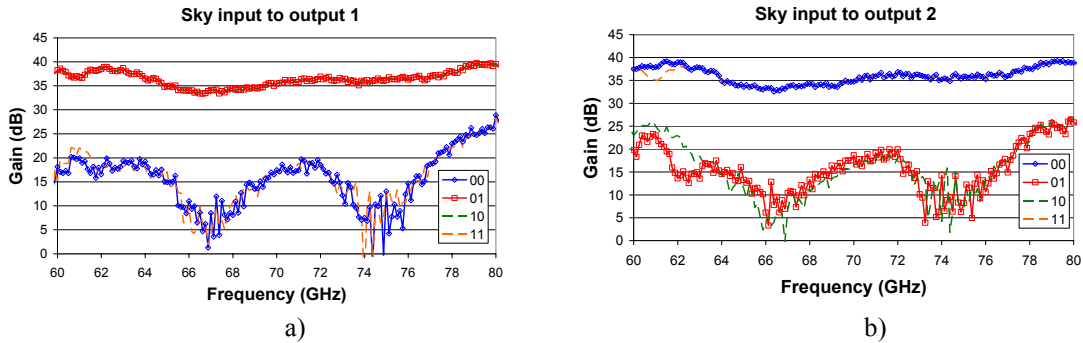

**Figure 12.** Gain and isolation measurements of a half FEM. Measurements were carried out for the four phase switch state combinations. Stimulus was fed to the sky input and measured a) from the first output and b) from the second output.

In each half FEM, there are two FEM_ACAs. With the two phase switches four possible combinations can be selected: 1) Both chains in the 0 degree phase switch state ('00'), 2) The first one in the 0° state and the second, in the 180° state ('01'), 3) The first one in the 180º and the second, in the 0º state ('10'), or 4) Both ones in the 180° state ('11'). Figure 12 a) shows measured gains of a half FEM (PFM 4) when stimulus from the VNA is fed into the sky input and measured from the first output. In two of the combinations ('01' and '10'), the sky input signal is passed to the first output. In the other two ('00' and '11'), the signal is isolated from the first output. In the '01' and '10' combinations, the gain profiles are practically identical in the required 63-77 GHz frequency range, as is the case when the other two combinations are compared. If the second output was measured instead (Figure 12 b), the combinations '01' and '10' are the ones isolated. When the signal is passed to an output, the gain is 35 dB or higher for almost the entire required range, and on average, the Planck requirement was fulfilled. If necessary, the BEM gain could also be increased slightly to ensure a high enough gain for the full receiver. In all six FEMs, the average channel gains ranged from 34.0-40.0 dB (uncertainty ±0.1 dB). When the signal is isolated from an output, the gain is 20 dB lower or more at all frequencies. This difference in gain is used as the measure for isolation. In all the six FEMs, the channel isolation values ranged from 11.3-22.1 dB (uncertainty ±0.1 dB).

**6.2 BEMs**

The BEM filter characteristics described in section 4 hold very accurately for every channel in the six BEMs. The -3 dB pass band, 62-81 GHz, was the same in every filter within 0.5 GHz. The BEM frequency response was measured as a function of input microwave power (from the VNA). Figure 13 a) shows the PFM2 BEM channel D responses. The overall shape of the responses in the pass band does not change as a function of the input power. Also, the pass bands roll at almost exactly 63 GHz and 77 GHz. The linearity of the channel is very good as well (figure 13 b), especially from -57 dBm upwards. The dynamic range was at least 15 dB from -57 dBm to -42 dBm. The responses were not characterised with higher input powers.

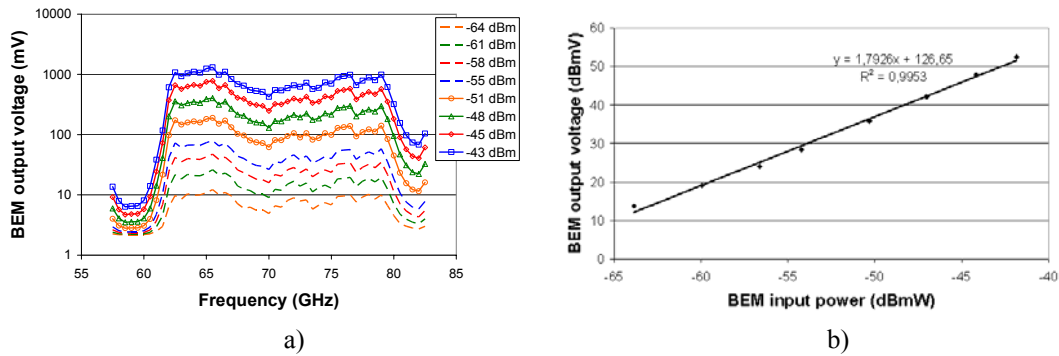

a)          b)

**Figure 13.** a) Measured Protoflight Model radiometer nr. 2 BEM channel D frequency responses at various input power levels (average power over the band is indicated). b) Channel D dynamic range at 70 GHz.

In three cases, the BEMs fulfilled the power consumption requirement, while the limit was exceeded for the other three. For the total six BEMs, the limit, 3.6 W, was exceeded by approximately 140 mW.

**6.3 FEM and BEM combined performance**

Figure 14 shows the results for the four whole channels of the PFM 4 radiometer (frequency step was 0.5 GHz). The channel effective bandwidths, $\Delta v_{eff}$, are defined here as the pre-detection bandwidths [15] and calculated numerically from

$$\Delta v_{eff} = \frac{\left[\int_0^\infty g(v)dv\right]^2}{\int_0^\infty g^2(v)dv}, \qquad (2.1)$$

where $v$ is the frequency and $g(v)$ the channel frequency response. The ripple was caused by the non-perfect matching of the sky load (TVA) output to the FEM input. In general, ripple in the band tend to narrow the ideal rectangular equivalent bandwidth. The effective bandwidths varied from 14.5-16.4 GHz. In all six radiometers, the range was from 10.3-16.4 GHz. The lowest value corresponds to PFM 6 radiometer, whose frequency response showed high ripple.

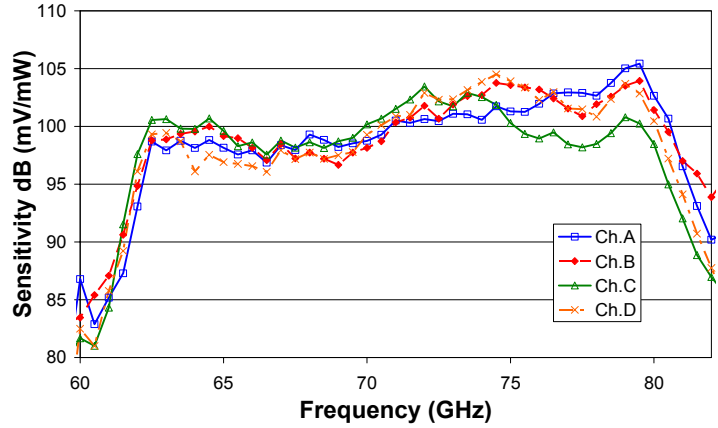

**Figure 14.** The measured Protoflight Model nr. 4 radiometer channel (A to D) sensitivities as a function of frequency.

Noise temperature is a measure for the total amount of noise due to amplifier chain. The noise temperature of the 70 GHz channels have to be as low as possible, especially because the CMB signal level is only about 2.7 K. The Planck requirement for the system noise temperature of a channel is less than or equal to 29.2 K. The noise temperatures were determined using the Y-factor technique. The linearity is very good in all channels in the 70 GHz radiometers. The noise temperature of the best channel of the six radiometers is 28.4 K, while the worst is 38.8 K. The Y-factor technique assumes that the sky loads and the reference loads are perfectly matched to the corresponding front-end module inputs; in practice the match is not perfect and causes an error in the measures here reported. Another factor of uncertainty is the load temperatures, because it is impossible to place the temperature sensors directly on the radiating surfaces. Determining the uncertainty of the noise temperature when using a complex test system, like in this case, requires Monte-Carlo methods. In case of the Elegant Breadboard [3, 4], where the same test setup was used as with the Protoflight Models, the uncertainty of a radiometer channel noise temperature was estimated to be ±5 K [16].

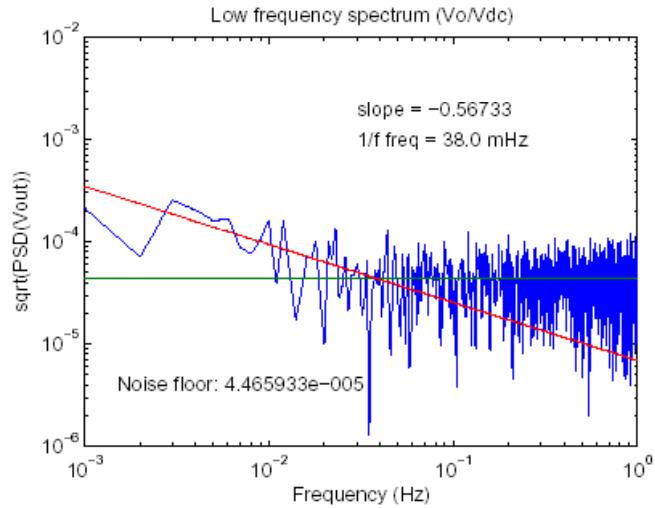

**Figure 15.** The measured Protoflight Model nr. 4 radiometer channel A 1/f noise spectrum density.

To test the stability of the receivers, hour-long time-series data were taken during normal radiometer operation from all output channels. Figure 15 shows an example of noise spectrum of the differentiated data from PFM 4. The two straight lines in figure 15 indicate the white spectrum noise level of the channel and the 1/f spectrum noise slope. The crossing point of the two lines gives the 1/f noise knee frequency, which is 38 mHz. Considering all values from the six radiometers the results are spread up to 248 mHz (worst case). Note that the 1/f noise measurements suffered from poor repeatability, which was due to external interference to the test environment. These results suggest that all radiometers were stable enough, but the test system, and particularly the sky load, was not. This was later confirmed by further tests performed for the integrated LFI instrument [17, 18].

Table 1 summarises the best, the worst and the average values of the key performance parameters. The shown uncertainties are based on worst case estimates as explained in the text previously.

**Table 1.** Summary of the 70 GHz Protoflight Model radiometer performance.

| Parameter | Requirement | PFM radiometers - best values | PFM radiometers - average values | PFM radiometers - worst values |
|---|---|---|---|---|
| FEM PERFORMANCE | | | | |
| FEM gain, dB | $\geq 35$ | 40.0±0.1 | 37.0±0.1 | 34.0±0.1 |
| FEM isolation, dB | $\geq 13$ | 22.1±0.1 | 18.5±0.1 | 11.3±0.1 |
| FEM power consumption, mW | $\leq 24$ | 21 | 23 | 25 |
| BEM PERFORMANCE | | | | |
| BEM filter pass band, GHz | 14 | 18.50±0.06 | 19.00±0.06 | 19.50±0.06 |
| BEM power consumption, mW | $\leq 604$ | 575 | 627 | 725 |
| RADIOMETER PERFORMANCE | | | | |
| System noise temperature, K | $\leq 29.2$ | 28±5 | 35±5 | 39±5 |
| White noise floor, $\times 10^{-5}$ V/$\sqrt{\text{Hz}}$ | | 1.5 | 2.8 | 4.8 |
| 1/f noise spectrum knee frequency, mHz | $\leq 50$ | 38 | 104 | 248 |
| Effective bandwidth, GHz | $\geq 14$ | 16 | 13 | 10 |

## 7. Conclusions

70 GHz Protoflight Model front-end and back-end modules were built for the ESA Planck Mission. Extensive tests and the measured performance have demonstrated that the present technology and the existing design can fulfil the mission requirements. The receiver is based on the InP MMIC low-noise amplifiers cooled to 20 K. The required stability has been obtained with continuous comparison receiver configuration. The best receiver performance are a noise temperature of 28.4 K, an effective bandwidth of 16.4 GHz and the 1/f noise knee frequency 38 mHz, which yield state-of-the-art performance at 70 GHz for the best receiver. On average, not all the receivers fully achieve the Planck goals, but they are still the most sensitive 70 GHz radiometers ever built for space applications. In general, we can conclude that the receiver meets the requirements of a compact size as well as low-power operation.

## Acknowledgments

Planck is a project of the European Space Agency (ESA) with instruments funded by ESA member states, and with special contributions from Denmark and the United States (NASA). The Planck-LFI project is developed by an international consortium led by Italy and involving Canada, Finland, Germany, Norway, Spain, Switzerland, UK, and USA.

The authors wish to thank the various funding agencies who have supported this work. In Finland, the Finnish Funding Agency for Technology and Innovation (Tekes), the Academy of Finland, the Waldemar von Frenckells stiftelse, the Magnus Ehrnrooth Foundation, and the Väisälä Foundation are gratefully acknowledged. In Italy, the Italian Space Agency for continuous support throughout the Planck Program is gratefully acknowledged. In the USA, the Planck project is supported by the NASA Science Mission Directorate.